# Predicting brain-age from raw T[1]-weighted Magnetic Resonance Imaging data using 3D Convolutional Neural Networks

Lukas Fisch[1†], Jan Ernsting[1,2], Nils R. Winter[1], Vincent Holstein[1], Ramona Leenings[1,2], Marie Beisemann[3], Kelvin Sarink[1], Daniel Emden[1], Nils Opel[1,4], Ronny Redlich[1,5], Jonathan Repple[1], Dominik Grotegerd[1], Susanne Meinert[1], Niklas Wulms[6], Heike Minnerup[6], Jochen G. Hirsch[7], Thoralf Niendorf[8], Beate Endemann[8], Fabian Bamberg[9], Thomas Kröncke[10], Annette Peters[11], Robin Bülow[12], Henry Völzke[13], Oyunbileg von Stackelberg[14], Ramona Felizitas Sowade[14], Lale Umutlu[15], Börge Schmidt[15], Svenja Caspers[16,17], German National Cohort Study Center Consortium, Harald Kugel[18], Bernhard T. Baune[1,19,20], Tilo Kircher[21], Benjamin Risse[2], Udo Dannlowski[1], Klaus Berger[6], Tim Hahn[1]

[1] Department of Psychiatry, University of Münster, Germany
[2] Department of Mathematics and Computer Science, University of Münster, Germany
[3] Department of Statistics, TU Dortmund University, Dortmund, Germany
[4] Interdisciplinary Center of Clinical Research (IZKF), University of Münster, Germany
[5] Department of Psychology, University of Halle, Halle, Germany
[6] Institute of Epidemiology and Social Medicine, University of Münster, Münster, Germany
[7] Fraunhofer MEVIS, Bremen, Germany
[8] Berlin Ultrahigh Field Facility (B.U.F.F.), NAKO imaging site Berlin, Max-Delbrueck Center for Molecular Medicine in the Helmholtz Association, Berlin, Germany
[9] Department of Radiology, Medical Center - University of Freiburg, Faculty of Medicine, University of Freiburg, Freiburg, Germany
[10] Department of Diagnostic and Interventional Radiology, University Hospital Augsburg, Augsburg, Germany
[11] Institute of Medical Information Processing, Biometry, and Epidemiology, Ludwig-Maximilians Universität, München, Germany
[12] Institute of Diagnostic Radiology and Neuroradiology, University of Greifswald, Greifswald, Germany
[13] Institute for Community Medicine, University Medicine Greifswald, Greifswald, Germany
[14] Department of Diagnostic and Interventional Radiology, University Hospital Heidelberg, Heidelberg, Germany; and Translational Lung Research Center, Member of the German Lung Research Center, Heidelberg, Germany
[15] Institute for Medical Informatics, Biometry and Epidemiology, University of Duisburg-Essen
[16] Institute for Anatomy I, Medical Faculty and University Hospital Düsseldorf, Heinrich Heine University Düsseldorf, 40225 Düsseldorf, Germany
[17] Institute of Neuroscience and Medicine (INM-1), Research Centre Jülich, 52425 Jülich, Germany
[18] University Clinic for Radiology, University of Münster, Germany
[19] Department of Psychiatry, Melbourne Medical School, The University of Melbourne, Melbourne, Australia
[20] The Florey Institute of Neuroscience and Mental Health, The University of Melbourne, Parkville, VIC, Australia
[21] Department of Psychiatry and Psychotherapy, Phillips University Marburg, Germany

[†]Corresponding author:

Lukas Fisch, Department of Psychiatry, University of Münster, Germany

Albert-Schweitzer-Campus 1, D-48149 Münster

Phone: +49 (0)2 51 / 83 – 56721, Fax:   +49 (0)2 51 / 83 – 56612, E-Mail: l.fisch@wwu.de

**Abstract**

Age prediction based on Magnetic Resonance Imaging (MRI) data of the brain is a biomarker to quantify the progress of brain diseases and aging. Current approaches rely on preparing the data with multiple preprocessing steps, such as registering voxels to a standardized brain atlas, which yields a significant computational overhead, hampers widespread usage and results in the predicted brain-age to be sensitive to preprocessing parameters. Here we describe a 3D Convolutional Neural Network (CNN) based on the ResNet architecture being trained on raw, non-registered $T_1$-weighted MRI data of N=10,691 samples from the German National Cohort and additionally applied and validated in N=2,173 samples from three independent studies using transfer learning. For comparison, state-of-the-art models using preprocessed neuroimaging data are trained and validated on the same samples. The 3D CNN using raw neuroimaging data predicts age with a mean average deviation of 2.84 years, outperforming the state-of-the-art brain-age models using preprocessed data. Since our approach is invariant to preprocessing software and parameter choices, it enables faster, more robust and more accurate brain-age modeling.

**1. Introduction**

The so-called *brain-age paradigm* aims to estimate the biological age (Ludwig and Smoke, 1980) of the brain of an individual under the assumption that brain-age may serve as a cumulative marker of disease-risk, functional capacity and residual lifespan (Cole, 2020). Specifically, brain-age models are trained to predict the chronological age of healthy subjects based on structural Magnetic Resonance Imaging (MRI) brain scans. Applying the trained model to brain scans of previously unseen individuals allows the determination of their brain age. A positive brain-age gap – i.e. a higher predicted brain-age compared to chronological age – has been associated with cognitive (Cole et al., 2018) and physiological aging (e.g. grip strength, lung function, walking speed), progression of dementia (Franke and Gaser, 2012; Gaser et al., 2013), mortality (Cole et al., 2018) and a range of neurological diseases and psychiatric disorders (for a review, see (Cole et al., 2019)).

Despite these insights and advances, the wide-spread use of current brain-age models is hampered by the need for extensive preprocessing. This regularly includes image registration, spatial smoothing and normalization to improve the quality and comparability of the data. Preprocessing thus manipulates the original image data e.g. via denoising, bias-correction and affine / non-linear transformations such that the voxels are registered to a standard space and unwanted artifacts are removed. This preprocessing leads to three issues: First, there is a tremendous computational overhead: For example, the preprocessing of the data used in this study took several weeks of CPU time. Second, if models are shared and applied to new datasets these must be preprocessed as well, which hampers widespread usage and clinical application. Third, preprocessing the same data with different software may yield substantially different results. Even seemingly minor updates of the same software tool – often unknown to the user – may affect the results. Likewise, if the preprocessing parameters are altered (e.g. because of different preferences between research institutes), all brain-age models may need to be retrained. Thus, preprocessing is a source for potential distortion of the brain-age predictions and a significant computational expense which reduces the useability of current brain-age models.

In the early years of brain-age research, efficient algorithms such as Support Vector Machines (SVM) and Relevance Vector Machines (RVM) have been used due to limited data and computation. When large datasets and increased computational performance became available, deep learning approaches such as Convolutional Neural Networks (CNNs) (LeCun et al., 1998) – which already constitute the state-of-the-art in neighboring fields including computer vision – have received increased attention in the brain-age community (Bashyam et al., 2020; Cole et al., 2017; Dinsdale et al., 2021; Hong et al., 2020). Weight sharing which is one main aspect of the CNN architecture enables feature extraction used to be done by the preprocessing to be integrated into the model while being able to fit into memory.

Because of the aforementioned disadvantages of preprocessing and the capabilities of CNNs reducing the preprocessing steps is a recent trend in the brain-age community (Bashyam et al., 2020; Cole et al., 2017; Dinsdale et al., 2021; Hong et al., 2020). In most articles, the preprocessing pipeline has been cut down to skull-stripping followed by voxel-based registration to an atlas. This reduced preprocessing yielded data which could be used by CNNs to model brain-age accurately. However, registration takes up a substantial part of the computation with regard to the preprocessing. Additionally, as pointed out in (Dinsdale et al., 2021) using nonlinearly registered images to train the CNN can lead to a model that is driven by artifacts of the registration process. One recent study attempted to model brain-age without any preprocessing in N = 220 children between the ages of 0 and 5 years (Hong et al., 2020). In addition to the low age range and low number of training samples, this approach results in the skull not being removed, potentially confounding brain age predictions with bone density and growth unrelated to actual brain aging.

We used an approach that reduced the preprocessing task to the minimum needed, which is skull-stripping. We hypothesize that this approach not only simplifies the use of brain-age models but will improve predictive performance due to less variability introduced by the preprocessing pipeline. The brain-age modeling using the skull-stripped data is done using a 3D CNN based on the ResNet architecture (He et al., 2016). The model is trained on the German National Cohort (GNC) brain MRI dataset (N = 10.691) which includes diverse samples regarding age, gender and scanner sites. To avoid performance overestimation (discussed in (Flint et al., 2019)) cross-validation and transfer learning to three different studies – i.e. BiDirect, FOR2107/MACS and IXI – is applied. The performance and computation times are compared to three state-of-the-art models trained with preprocessed data from the conventional preprocessing pipeline (Hahn et al., 2020). Finally, a bias assessment is done to investigate the influence of gender and ethnicity on the brain-age predictions.

## 2. Materials and Methods

### 2.1 Datasets

<u>Training and Validation Samples</u>

This study utilizes existing data from the German National Cohort (GNC: https://nako.de/), the BiDirect study (https://www.medizin.uni-muenster.de/en/epi/research/projects/bidirect.html), the Marburg-Münster-Affective-Cohort Study (FOR2107/MACS: https://for2107.de/), the Information eXtraction from Images (IXI) (https://brain-development.org/ixi-dataset/) and the Beijing Normal University dataset (http://fcon_1000.projects.nitrc.org/indi/retro/BeijingEnhanced.html). Data availability is governed by the respective consortia. No new data was acquired for this study.

The models are trained on the GNC dataset using 10-fold cross-validation. Additionally, the models are trained on the total GNC dataset and applied to the BiDirect, FOR2107/MACS, and IXI datasets using transfer learning. Ethnicity bias assessment is conducted using the Beijing Normal University dataset. In the following, each dataset is described in detail. Also, Table 1 provides further sample characteristics, including sample sizes, sex distribution, age minimum and maximum as well as standard deviation.

*German National Cohort (GNC):* This cohort is a large-scale population study which examined 205.000 subjects, aged 20 to 72 years in 18 study centers across Germany between 2014 and 2019 using a comprehensive program. The study included a 3.0 Tesla whole-body MRI examination involving a $T_1$-weighed whole brain scan ($T_1$w-MPRAGE) in 30.000 participants, performed in five GNC imaging centers equipped with dedicated identical MR systems (Skyra, Siemens Healthineers, Erlangen, Germany) examining participants from 11 of the centers. Our analysis is based on the 'data freeze 100K' milestone for the first 100.000 participants which also included the first 10,691 participants with completed MRIs of sufficient quality (for the detailed MRI protocol, see (Ahrens et al., 2014; Bamberg et al., 2015)).

*BiDirect:* The BiDirect Study is a propspective project that comprises three distinct cohorts: patients hospitalized for an acute episode of major depression, patients up to six months after an acute cardiac event and healthy controls randomly drawn from the population register of the city of Münster, Germany. Baseline examination of all participants included a 3 Tesla MRI of the brain, a computer-assisted face-to-face interview on sociodemographic characteristics, medical history, and an extensive psychiatric characterization as well as a collection of blood samples. Inclusion criteria for the present analyses were availability of completed baseline MRI data and sufficient MRI quality. Only data from the healthy control group was used in this study. Further details on the rationale, design and recruitment procedures of the BiDirect study have been described in (Teuber et al., 2017).

*Marburg-Münster Affective Disorders Cohort Study (FOR2107/MACS):* Participants – patients and a control group - were recruited through psychiatric hospitals or newspaper advertisements. Inclusion criteria included mild, moderate or partially remitted Major Depressive Disorder episodes in addition to severe depression. Patients could be undergoing inpatient, outpatient or no current treatment. The FOR2107/MACS was conducted at two scanning sites – University of Münster and University of Marburg. Further details about the structure of the FOR2107/MACS (Kircher et al., 2019) and MRI quality assurance protocol (Vogelbacher et al., 2018) are provided elsewhere. Inclusion criteria for the present study were availability of completed baseline MRI data with sufficient MRI quality. Only data from the healthy control group was used in this study.

*Information eXtraction from Images (IXI):* This dataset comprises images from healthy subjects, along with demographic characteristics, collected as part of the Information eXtraction from Images (IXI) project available for download (https://brain-development.org/ixi-dataset/). The data has been collected at three different hospitals in London (Hammersmith Hospital using a Philips 3T system), Guy's Hospital using a Philips 1.5T system, and Institute of Psychiatry using a GE 1.5T system). Inclusion criteria for the present study were availability of completed baseline MRI data with sufficient MRI quality.

*Beijing Normal University:* This dataset includes 180 healthy controls from a community (student) sample at Beijing Normal University in China. Inclusion criteria for the present study were availability of completed baseline MRI data with sufficient MRI quality. Further details can be found online (http://fcon_1000.projects.nitrc.org/indi/retro/BeijingEnhanced.html).

Table 1. Overview of sample and subsample distributions.

| Sample | Group | N | N Males | Age Mean | Age Std. | Age Min. | Age Max. |
|---|---|---|---|---|---|---|---|
| GNC | Population Sample | 10,691 | 5,485 | 51.79 | 11.37 | 20 | 72 |
| BiDirect | Population Sample | 688 | 361 | 53.01 | 8.48 | 35 | 70 |
| FOR2107/MACS | HC | 924 | 595 | 34.16 | 12.88 | 18 | 65 |
| IXI | HC | 561 | 311 | 48.62 | 16.49 | 20 | 86 |
| Beijing Normal University | HC | 179 | 107 | 21.25 | 1.92 | 18 | 28 |

Std.: Standard Deviation, Min.: Minimum, Max.: Maximum, HC: Healthy Controls

## 2.2 Preprocessing

In order to compare our approach to other models using preprocessing, we require the same data in both conventionally preprocessed and minimally preprocessed formats. For both formats, we additionally consider three different spatial resolutions to investigate its influence on model performances. The highest spatial resolution considered is 134 x 168 x 134 voxels, resulting in a voxel size of 1.36 x 1.30 x 1.36 mm, because this is found to be the average cuboid of voxels which is taken up by the brain in the $T_1$-weighted images in the GNC dataset. The other considered resolutions are 68 x 84 x 68 and 34 x 42 x 34 voxels, yielding a voxel size of 2.73 x 2.63 x 2.73 mm and 5.46 x 5.27 x 5.46 mm, respectively.

For minimal preprocessing, the $T_1$-weighted images only ran through skull stripping, cropping and resampling. The skull stripping is done with the HD-BET brain extraction tool (Isensee et al., 2019) which uses an artificial neural network with an U-Net architecture (Ronneberger et al., 2015) which is specialized for image segmentation tasks. After the skull is removed the brain is cropped and finally this cropped part of the scan is resampled via linear interpolation to the three considered spatial resolutions without registration to a standard space. The minimal preprocessing is done using the implementations of HD-BET (https://github.com/MIC-DKFZ/HD-BET) and fastai_scans (https://github.com/renato145/fastai_scans).

For the processed data, the $T_1$-weighted images are preprocessed with the commonly used CAT12 toolbox (built 1450 with SPM12 version 7487 and Matlab 2019a; http://dbm.neuro.uni-jena.de/cat) with default parameters. This preprocessing includes a bias-correction, a tissue segmentation, a DARTEL-normalization to MNI-space, smoothening with a Gaussian Kernel (8mm FWHM) and a resampling to the three considered voxel sizes. The application of these preprocessing steps to the GNC dataset took several weeks of CPU time. After this first part of the preprocessing, a whole-brain mask comprising all grey-matter voxels is applied, data is vectorized, features with zero-variance in the GNC dataset are removed, and the scikit-learn (Pedregosa et al., 2011) Standard Scaler is applied.

## 2.3 Data Augmentation

Data Augmentation artificially increases the amount of training data which avoids overfitting and yields increased stability of the model. In this paper the data is augmented by cropping the three-dimensional brain-scan (see Figure 1). The size of the cropped image along each dimension ranges from 90% to 100% of the original image and is randomly picked for each dimension individually. After cropping at a random position, oversampling is done via linear interpolation such that the original resolution of the skull stripped $T_1$-scan is restored.

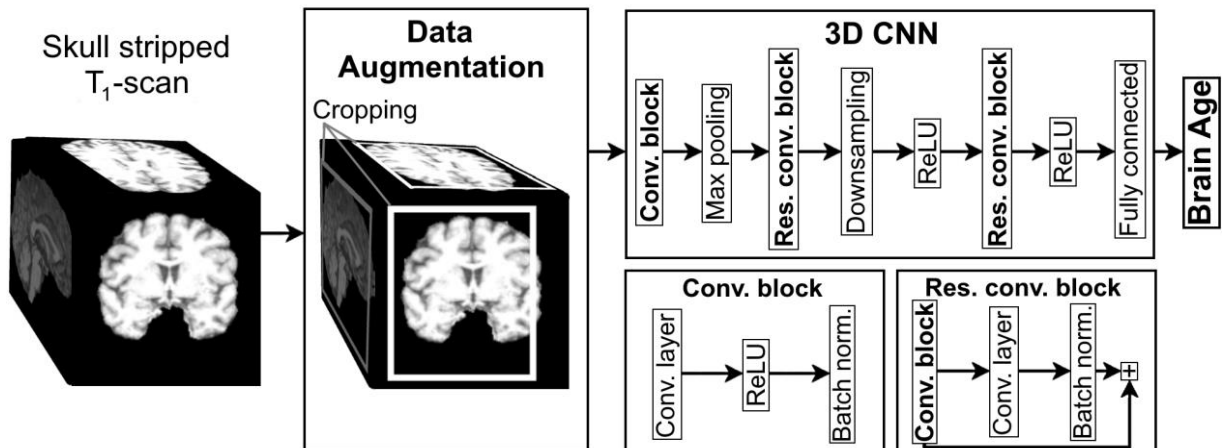

3D CNN: 3D Convolutional Neural Network, Conv.: Convolutional, Res.: Residual, Norm.: Normalization, ReLU: Rectified Linear Unit

Figure 1. Schematic representation of the machine learning pipeline used to predict brain-age from raw, non-registered, skull stripped $T_1$-scans. White rectangles indicate the cropping of the cubic scan before the 3D CNN with the ResNet architecture is applied.

## 2.4 Model Architecture

*3D CNN (raw data):* The model which predicts brain-age from minimally preprocessed data is a 3D convolutional neural network (CNN) with the ResNet architecture (He et al., 2016) (see Figure 1). Two blocks, the convolutional and the residual convolutional block, are used repeatedly in this architecture. The convolutional block starts with a convolutional layer which is followed by a rectified linear unit (ReLU) and a 3D batch-normalization layer (Ioffe and Szegedy, 2015). The first convolutional block in the model uses a kernel size of 7 x 7 x 7, a stride of 2 x 2 x 2, a padding of 3 x 3 x 3 and 16 output channels. All the following convolutional blocks which are part of residual convolutional blocks take 16 respectively 32 input channels and use a kernel size of 3 x 3 x 3, a stride of 1 x 1 x 1, a padding of 1 x 1 x 1 and 32 output channels. The first module of the residual convolutional block is the just described convolutional block followed by a convolutional layer and a 3D batch-normalization layer.

The proposed model consists of a convolutional block followed by two residual convolutional blocks. After the convolutional block max pooling is applied with a kernel size of 3 x 3 x 3, a stride of 2 x 2 x 2 and a padding of 1 x 1 x 1. The first residual convolutional block is followed by a convolutional layer, a 3D batch-normalization layer and a ReLU is applied. The second residual convolutional block is activated by a ReLU and followed by the final module which consists of a 1D batch-normalization layer and a dense net with a sigmoid activation function. The 3D CNN using the high resolution of 136 x 168 x 136 voxels consists of 49,820,497 parameters. The corresponding models using medium (68 x 84 x 68 voxels) and low resolution (34 x 42 x 34 voxels) consist of 6,317,905 and 1,015,633 parameters, respectively.

*ANN (preprocessed data):* The artificial neural net (ANN) used in this work is fully connected consisting of one hidden layer with 32 nodes. These nodes are activated by a ReLU which is followed by a 1D batch-normalization and a dropout layer (Srivastava et al., 2014).

*SVM (preprocessed data):* In this paper the sklearn implementation of the Support Vector Regression (SVR) with default settings and a linear kernel is used (Pedregosa et al., 2011).

*RVM (preprocessed data):* The application of the Relevance Vector Regression (RVR) is done using the fast-rvm implementation from sklearn-bayes (https://github.com/AmazaspShumik/sklearn-bayes) with default settings.

## 2.5 Training Procedure

The 3D CNN is trained using the Adam W algorithm to minimize the mean squared error (MSE) using the fastai package (Howard and Gugger, 2020; Kingma and Ba, 2015). Part of this package is the 1-cycle learning rate policy which maximizes convergence by increasing the learning rate to a maximum $\eta_{max}$ during the begin of the training and decreasing it gradually afterwards (Smith, 2018). Before training, an optimal maximal learning rate $\eta_{max}$ is determined via a mock training during which the learning rate is increased exponentially. As proposed in (Howard and Gugger, 2020), a tenth of the learning rate with the minimal loss is used as the optimal maximal learning rate $\eta_{max}$. This yields an optimal maximal learning rate of $\eta_{max}$ = .02. The minimum and maximum value of the cyclical momentums used in the 1-cycle method are set to 0.85 and 0.95.

The minimization of the MSE for the ANN using the preprocessed data is done by using the Adam algorithm (Kingma and Ba, 2015) with a learning rate of .01 as proposed in (Hahn et al., 2020). Both the SVR and the RVR are trained using the default fitting method of their respective implementations.

**3. Results**

The minimal preprocessing reduces the time needed for data preparation by an order of magnitude. The application of the conventional, CPU-only preprocessing to the GNC dataset takes two weeks on the considered system, a personal computer with a Ryzen 9 5950X Central Processing Unit (CPU) and a RTX 3090 Graphics Processing Unit (GPU). In contrast the GPU supported minimal preprocessing of the same dataset can be done in one day on the same system.

On top of that, the 3D CNN using these minimally preprocessed brain scans with the highest spatial resolution considered (134 x 168 x 134 voxels) achieves the best average mean absolute error (MAE) of 2.84 years in the validation folds during 10-fold cross validation with a standard deviation (std) of 0.05 years (see Table 1). This is a significant improvement (p<.001) compared to the ANN with a MAE of 3.32 years (std=.05) which is the best results of the considered state-of-the-art models using conventionally preprocessed scans. The SVR could not be trained with the highest resolution data because this would have been needed >100 GB of memory. RVR could be computed with high resolution performing with the highest MAE of 4.34 (std=.11).

The 3D CNN also performs with the lowest MAEs for both the medium (68 x 84 x 68 voxels) and the low resolution (34 x 42 x 34 voxels) of 3.13 years (std=.07) and 3.75 years (std=.08), respectively. These larger voxel size reduces the time needed to train with the total GNC dataset from 285 minutes down to 35 (medium resolution) and 10 minutes (low resolution), respectively. The much simpler ANN which needs only 14 minutes to train on the high resolution data can be trained in two minutes on medium resolution and in under one minute on low resolution scans. However, these shorter training times have to be put in context with the additional time needed for conventional preprocessing. On top of that, the MAEs of the ANN are 3.38 (std=.10) and 4.04 years (std=.05) which is worse than the 3D CNNs results. The SVR and the RVR result in the highest MAEs of over four years up to 6.88 years (std=.13) using the SVR on the low resolution scans. In this case, the optimizer does not converge after the maximum of iterations thereby causing the high MAE and 13 minutes of training time. The RVR can be trained quickly with training times from two minutes (high resolution) down to ten seconds (low resolution), however yielding high MAEs.

Table 2. Mean Absolute Error (MAE) and training runtime for all considered models and scan resolutions using the German National Cohort (GNC) dataset. The standard deviation of the MAE across folds is given in parentheses. The training runtime is the time needed to train the model once using all N = 10.691 samples as training data on our system (CPU: Ryzen 9 5950X, GPU: RTX 3090).

| Metric | Mean Absolute Error | | | Training runtime [minutes] | | |
|---|---|---|---|---|---|---|
| Resolution | High | Medium | Low | High | Medium | Low |
| RVR<br>*conv. proc.* | 4.34 (.11) | 4.43 (.12) | 4.58 (.10) | 2 | 1 | <1 |
| SVR<br>*conv. proc.* | - | 4.16 (1.31) | 6.88 (.13) | - | 2 | 13 |
| ANN<br>*conv. proc.* | 3.22 (.06) | 3.38 (.10) | 4.04 (.05) | 14 | 2 | <1 |
| 3D-CNN<br>*min. proc.* | **2.84 (.05)** | **3.13 (.07)** | **3.75 (.08)** | 285 | 35 | 10 |

RVR: Relevance Vector Regression, SVR: Support Vector Regression, ANN: Artificial Neural Network
3D CNN: 3D Convolutional Neural Network, Conv. proc.: Conventional Preprocessing, Min. proc.: Minimal preprocessing

To check how well the models can be transferred to other datasets, the models are applied to three independent datasets (N=2,173), namely the BiDirect study, the FOR2107/MACS, and the IXI dataset (IXI) which cover a larger age range than the training data (20 to 86 years vs. 20 to 72 years in the GNC sample) and were acquired on independent scanners and sites. This application shows how additional sources of variability such as different data acquisition protocols, alterations in recruitment or sample characteristics can be learned starting from the model trained on the full GNC dataset. In addition to simply applying the pretrained model to these datasets, transfer learning is done on the respective dataset with the same training routine which is used during the initial training with the GNC dataset.

Table 3. Mean Absolute Error for all considered scan resolutions and models which were pretrained on the German National Cohort (GNC) dataset and successively applied to three independent datasets (BiDirect, FOR2107/MACS, and IXI). *TL*: Models which did undergo transfer learning on respective datasets after pretraining. For both the TL-models, the standard deviation of the MAE during cross-validation is given in parentheses.

| Study | BiDirect | | | FOR2107/MACS | | | IXI | | |
|---|---|---|---|---|---|---|---|---|---|
| **Resolution** | **High** | **Medium** | **Low** | **High** | **Medium** | **Low** | **High** | **Medium** | **Low** |
| RVR *conv. proc.* | 4.65 | 4.66 | 4.8 | 6.09 | 6.28 | 6.72 | 6.27 | 6.37 | 6.55 |
| SVR *conv. proc.* | - | 5.59 | 7.21 | - | 6.36 | 7.77 | - | 7.41 | 8.19 |
| ANN *conv. proc.* | 4.03 | 4.17 | 4.91 | 4.59 | 4.98 | 5.61 | 5.81 | 5.62 | 6.44 |
| 3D CNN *min. proc.* | 11.84 | 8.02 | 9.61 | 5.75 | 7.32 | 8.16 | 8.25 | 7.85 | 8.88 |
| ANN (TL) *conv. proc.* | 3.63 (.31) | 3.63 (.31) | 4.31 (.31) | 3.49 (.24) | 3.59 (.22) | 4.25 (.36) | 3.77 (.36) | 3.71 (.29) | 4.47 (.52) |
| 3D CNN (TL) *min. proc.* | 3.34 (.20) | 3.51 (.24) | 4.41 (.26) | 2.96 (.15) | 3.40 (.23) | 4.37 (.37) | 3.77 (.49) | 4.15 (.47) | 5.09 (.60) |

RVR: Relevance Vector Regression, SVR: Support Vector Regression, ANN: Artificial Neural Network, 3D CNN: 3D Convolutional Neural Network, Conv. proc.: Conventional Preprocessing, Min. proc.: Minimal preprocessing, TL: Transfer Learning

Without transfer learning the 3D CNN does perform worst across datasets and used resolutions. Only in case of the FOR2107/MACS dataset the 3D CNN yields a competitive MAE of 5.75 years, which is better than the RVRs 6.09 years but still worse than the ANNs 4.59 years. The ANN using the preprocessed data shows to predict age most robustly across studies reaching a MAE of as low as 4.03 years for the BiDirect study.

However, when transfer learning is utilized the 3D CNN again yields the lowest MAEs for each study. These low MAEs are always produced by the 3D CNN using the highest resolution scans. The 3D CNNs performance drops faster when compared to the ANN when the resolution of the scans is reduced. In both the BiDirect (N=688) and the FOR2107/MACS (N=924) (Vogelbacher et al., 2018) dataset the 3D CNN still reaches the lowest MAEs of 3.51 (std=.24) and 3.40 years (std=.23) using the medium resolution. But in case of the IXI dataset (N=561) a MAE of 4.15 years (std=.47) is reached which favors the ANN with an MAE of 3.71 years (std=.29). In case of the low resolution the ANN using preprocessed data results in lower MAEs in all the three datasets considered.

<u>Bias Assessment</u>

As machine learning models are not programmed, but trained, they will mimic systematic biases inherent in their training data (Cearns et al., 2019). While this potential algorithmic bias must be carefully investigated with regard to performance differences in specific subgroups to determine for which populations it yields robust estimates, it is often neglected in brain-age research (for an in-depth discussion, see (Sokol et al., 2019)). Here, the model performance differences for sex, ethnicity, and age for the 3D CNN are assessed, using the high resolution of 136 x 168 x 136 voxels which yielded the best results across studies. Due to different age ranges between men and women, the standardized MAE is used as a performance metric.

The accuracy of the predicted ages is not substantially dependent on the sex of the samples reaching a standardized MAE of .388 for women and a standardized MAE of .386 for males in the BiDirect dataset. In the MACS dataset, standardized MAE is .233 in women and .227 in men. For the IXI dataset, the 3D CNN reaches a standardized MAE of .215 in women and .242 in men. Note that sex here is defined as biologically male and female as no further information on gender was available.

Regarding ethnicity, the 3D CNN is tested on the publicly available dataset from Beijing Normal University (N=179) with the subjects being between 18 and 28 years of age. The 3D CNN trained on this dataset in the same manner as the BiDirect, FOR2107/MACS and IXI dataset using transfer learning yielded a MAE of 1.44 years (std=.33) and a standardized MAE of .750. This substantially higher standardized MAE indicates that the model is affected by the ethnicity of the subjects. Finally, the model performance across age is investigated. As commonly reported in brain-age studies, the brain-age gap and the age are correlated with r=-.30 in the GNC, r=-.24 in the BiDirect, r=-.35 in the FOR2107/MACS and r=-.38 for the IXI dataset.

**4. Discussion**

We showed that using minimally preprocessed brain scans in combination with a 3D CNN improves predictive performance compared to state-of-the-art brain-age models which use preprocessed neuroimaging data. The 3D CNN achieved a MAE of 2.84 years (std=.05) in the GNC dataset including 10,691 men and women aged 20 – 72 years. This is a significant improvement compared to the RVR, SVR and ANN model which were trained on preprocessed data and whose best performance is a MAE of 3.22 years (std=.06) achieved by the ANN. The results after transfer learning on three independent datasets confirm the advantage of the 3D CNN in terms of being more adaptive to unseen datasets. However, when the pretrained model is applied to these datasets without transfer learning the models using the preprocessed data show more robust age predictions. This issue could be reduced by incorporating additional data augmentation methods, which aim to match artifacts of different MRI machines such as different noise levels.

Assessment of the model bias showed that the performance of the 3D CNN does not differ substantially between sexes. However, there are biases related to ethnicity and age groups which could be uncovered during the assessment. Transfer learning on the dataset consisting solely of young Chinese citizens showed a much higher standardized MAE. The correlation between BAG and age is a known issue in brain-age research, as the predictions after regression will tend towards the mean of the dataset. To counteract this bias while using these brain-age predictions in further analysis, age should be included as a covariate (for an introduction to age-bias correction approaches in brain-age modeling, see (de Lange and Cole, 2020)).

Regarding the age bias, the model could be improved by using the methods presented in (Smith et al., 2019) that reduced the tendency of the age-prediction towards the mean of the dataset. The ethnicity bias could be mitigated by finetuning the model using datasets with heterogeneous ethnicity. Finally, uncertainty of the brain-age prediction could be estimated using this 3D CNN by pursuing the methods shown in (Hahn et al., 2020).

The results confirm that the proposed 3D CNN brain-age research can be done without spending days or weeks on extensive preprocessing of the neuroimaging data. A minimal preprocessing consisting of HD-BET for brain-extraction and cropping to minimize the amount of zero intensity voxels, appears to be sufficient. This minimal preprocessing can be applied to large datasets within one day using a single GPU. Additionally, resampling to lower resolutions can be used to train models on large datasets within minutes while still obtaining reasonable performances and for hyperparameter finetuning of the model. The implementation of the minimal preprocessing, the training routine and the pretrained models is made available at https://github.com/wwu-mmll/fastbrainage. This user-friendly implementation aims to further lowering the barrier for researchers aiming to apply the model to their own dataset.

Funding

This work was funded by the German Research Foundation (DFG grants HA7070/2-2, HA7070/3, HA7070/4 to TH) and the Interdisciplinary Center for Clinical Research (IZKF) of the medical faculty of Münster (grants Dan3/012/17 to UD and MzH 3/020/20 to TH).

The analysis was conducted with data from the German National Cohort (GNC) (www.nako.de). The GNC is funded by the Federal Ministry of Education and Research (BMBF) [project funding reference numbers: 01ER1301A/B/C and 01ER1511D], the federal states and the Helmholtz Association with additional financial support by the participating universities and the institutes of the Helmholtz Association and of the Leibniz Association.  We thank all participants who took part in the GNC study and the staff in this research program.

The BiDirect Study is supported by grants from the Federal Ministry of Education and Research (BMBF) to the University of Muenster (grant numbers: 01ER0816, 01ER1506 and 01ER1205).

The MACS dataset used in this work is part of the German multicenter consortium "Neurobiology of Affective Disorders. A translational perspective on brain structure and function", funded by the German Research Foundation (Deutsche Forschungsgemeinschaft DFG; Forschungsgruppe/Research Unit FOR2107). Principal investigators (PIs) with respective areas of responsibility in the FOR2107 consortium are: Work Package WP1, FOR2107/MACS cohort and brainimaging: Tilo Kircher (speaker FOR2107; DFG grant numbers KI 588/14-1, KI 588/14-2), Udo Dannlowski (co-speaker FOR2107; DA 1151/5-1, DA 1151/5-2), Axel Krug (KR 3822/5-1, KR 3822/7-2), Igor Nenadic (NE 2254/1-2), Carsten Konrad (KO 4291/3-1). WP2, animal phenotyping: Markus Wöhr (WO 1732/4-1, WO 1732/4-2), Rainer Schwarting (SCHW 559/14-1, SCHW 559/14-2). WP3, miRNA: Gerhard Schratt (SCHR 1136/3-1, 1136/3-2). WP4, immunology, mitochondriae: Judith Alferink (AL 1145/5-2), Carsten Culmsee (CU 43/9-1, CU 43/9-2), Holger Garn (GA 545/5-1, GA 545/7-2). WP5, genetics: Marcella Rietschel (RI 908/11-1, RI 908/11-2), Markus Nöthen (NO 246/10-1, NO 246/10-2), Stephanie Witt (WI 3439/3-1, WI 3439/3-2). WP6, multi method data analytics: Andreas Jansen (JA 1890/7-1, JA 1890/7-2), Tim Hahn (HA 7070/2-2), Bertram Müller-Myhsok (MU1315/8-2), Astrid Dempfle (DE 1614/3-1, DE 1614/3-2). CP1, biobank: Petra Pfefferle (PF 784/1-1, PF 784/1-2), Harald Renz (RE 737/20-1, 737/20-2). CP2, administration. Tilo Kircher (KI 588/15-1, KI 588/17-1), Udo Dannlowski (DA 1151/6-1), Carsten Konrad (KO 4291/4-1). Data access and responsibility: All PIs take responsibility for the integrity of the respective study data and their components. All authors and coauthors had full access to all study data. The FOR2107 cohort project (WP1) was approved by the Ethics Committees of the Medical Faculties, University of Marburg (AZ: 07/14) and University of Münster (AZ: 2014-422-b-S).

Financial support for the Beijing Normal University dataset used in this project was provided by a grant from the National Natural Science Foundation of China: 30770594 and a grant from the National High Technology Program of China (863): 2008AA02Z405.

# References

Ahrens, W., Hoffmann, W., Jöckel, K.H., Kaaks, R., Gromer, B., Greiser, K.H., Linseisen, J., Schmidt, B., Wichmann, H.E., Weg-Remers, S., 2014. The German National Cohort: Aims, study des. Eur. J. Epidemiol. 29, 371–382. https://doi.org/10.1007/s10654-014-9890-7

Bamberg, F., Kauczor, H.U., Weckbach, S., Schlett, C.L., Forsting, M., Ladd, S.C., Greiser, K.H., Weber, M.A., Schulz-Menger, J., Niendorf, T., Pischon, T., Caspers, S., Amunts, K., Berger, K., Bülow, R., Hosten, N., Hegenscheid, K., Kröncke, T., Linseisen, J., Günther, M., Hirsch, J.G., Köhn, A., Hendel, T., Wichmann, H.E., Schmidt, B., Jöckel, K.H., Hoffmann, W., Kaaks, R., Reiser, M.F., Völzke, H., 2015. Whole-body MR imaging in the German national cohort: Rationale, design, and technical background1. Radiology. https://doi.org/10.1148/radiol.2015142272

Bashyam, V.M., Erus, G., Doshi, J., Habes, M., Nasralah, I., Truelove-Hill, M., Srinivasan, D., Mamourian, L., Pomponio, R., Fan, Y., Launer, L.J., Masters, C.L., Maruff, P., Zhuo, C., Völzke, H., Johnson, S.C., Fripp, J., Koutsouleris, N., Satterthwaite, T.D., Wolf, D., Gur, R.E., Gur, R.C., Morris, J., Albert, M.S., Grabe, H.J., Resnick, S., Bryan, R.N., Wolk, D.A., Shou, H., Davatzikos, C., 2020. MRI signatures of brain age and disease over the lifespan based on a deep brain network and 14 468 individuals worldwide. Brain 143, 2312–2324. https://doi.org/10.1093/brain/awaa160

Cearns, M., Hahn, T., Baune, B.T., 2019. Recommendations and future directions for supervised machine learning in psychiatry. Transl. Psychiatry 9, 271. https://doi.org/10.1038/s41398-019-0607-2

Cole, J.H., 2020. Multimodality neuroimaging brain-age in UK biobank: relationship to biomedical, lifestyle, and cognitive factors. Neurobiol. Aging 92, 34–42. https://doi.org/10.1016/j.neurobiolaging.2020.03.014

Cole, J.H., Marioni, R.E., Harris, S.E., Deary, I.J., 2019. Brain age and other bodily 'ages': implications for neuropsychiatry. Mol. Psychiatry 24, 266–281. https://doi.org/10.1038/s41380-018-0098-1

Cole, J.H., Poudel, R.P.K., Tsagkrasoulis, D., Caan, M.W.A., Steves, C., Spector, T.D., Montana, G., 2017. Predicting brain age with deep learning from raw imaging data results in a reliable and heritable biomarker. Neuroimage 163, 115–124. https://doi.org/10.1016/j.neuroimage.2017.07.059

Cole, J.H., Ritchie, S.J., Bastin, M.E., Valdés Hernández, M.C., Muñoz Maniega, S., Royle, N., Corley, J., Pattie, A., Harris, S.E., Zhang, Q., Wray, N.R., Redmond, P., Marioni, R.E., Starr, J.M., Cox, S.R.,

Wardlaw, J.M., Sharp, D.J., Deary, I.J., 2018. Brain age predicts mortality. Mol. Psychiatry 23, 1385–1392. https://doi.org/10.1038/mp.2017.62

de Lange, A.M.G., Cole, J.H., 2020. Commentary: Correction procedures in brain-age prediction. NeuroImage Clin. 26. https://doi.org/10.1016/j.nicl.2020.102229

Dinsdale, N.K., Bluemke, E., Smith, S.M., Arya, Z., Vidaurre, D., Jenkinson, M., Namburete, A.I.L., 2021. Learning patterns of the ageing brain in MRI using deep convolutional networks. Neuroimage 224. https://doi.org/10.1016/j.neuroimage.2020.117401

Flint, C., Cearns, M., Opel, N., Redlich, R., Mehler, D.M.A., Emden, D., Winter, N.R., Leenings, R., Eickhoff, S.B., Kircher, T., Krug, A., Nenadic, I., Arolt, V., Clark, S., Baune, B.T., Jiang, X., Dannlowski, U., Hahn, T., 2019. Systematic Overestimation of Machine Learning Performance in Neuroimaging Studies of Depression.

Franke, K., Gaser, C., 2012. Longitudinal changes in individual BrainAGE in healthy aging, mild cognitive impairment, and Alzheimer's Disease. GeroPsych J. Gerontopsychology Geriatr. Psychiatry 25, 235–245. https://doi.org/10.1024/1662-9647/a000074

Gaser, C., Franke, K., Klöppel, S., Koutsouleris, N., Sauer, H., 2013. BrainAGE in Mild Cognitive Impaired Patients: Predicting the Conversion to Alzheimer's Disease. PLoS One 8, e67346. https://doi.org/10.1371/journal.pone.0067346

Hahn, T., Ernsting, J., Winter, N.R., Holstein, V., Leenings, R., Beisemann, M., Fisch, L., Sarink, K., Emden, D., Opel, N., Redlich, R., Repple, J., Grotegerd, D., Meinert, S., Kugel, H., Baune, B.T., Kircher, T., Gaser, C., Cole, J.H., Dannlowski, U., Berger, K., 2020. An Uncertainty-Aware, Shareable and Transparent Neural Network Architecture for Brain-Age Modeling.

He, K., Zhang, X., Ren, S., Sun, J., 2016. Deep residual learning for image recognition. Proc. IEEE Comput. Soc. Conf. Comput. Vis. Pattern Recognit. 2016-Decem, 770–778. https://doi.org/10.1109/CVPR.2016.90

Hong, J., Feng, Z., Wang, S.H., Peet, A., Zhang, Y.D., Sun, Y., Yang, M., 2020. Brain Age Prediction of Children Using Routine Brain MR Images via Deep Learning. Front. Neurol. 11, 584682. https://doi.org/10.3389/fneur.2020.584682

Howard, J., Gugger, S., 2020. Fastai: A layered api for deep learning. Inf. 11, 108. https://doi.org/10.3390/info11020108

Ioffe, S., Szegedy, C., 2015. Batch normalization: Accelerating deep network training by reducing internal covariate shift. 32nd Int. Conf. Mach. Learn. ICML 2015 1, 448–456.

Isensee, F., Schell, M., Tursunova, I., Brugnara, G., Bonekamp, D., Neuberger, U., Wick, A., Schlemmer, H., Heiland, S., Wick, W., Bendszus, M., Maier-Hein, K.H., Kickingereder, P., 2019. Automated brain extraction of multi-sequence MRI using artificial neural networks. Hum. Brain Mapp. 40, 1–13.

Kingma, D.P., Ba, J.L., 2015. Adam: A method for stochastic optimization. 3rd Int. Conf. Learn. Represent. ICLR 2015 - Conf. Track Proc.

Kircher, T., Wöhr, M., Nenadic, I., Schwarting, R., Schratt, G., Alferink, J., Culmsee, C., Garn, H., Hahn, T., Müller-Myhsok, B., Dempfle, A., Hahmann, M., Jansen, A., Pfefferle, P., Renz, H., Rietschel, M., Witt, S.H., Nöthen, M., Krug, A., Dannlowski, U., 2019. Neurobiology of the major psychoses: a translational perspective on brain structure and function—the FOR2107 consortium. Eur. Arch. Psychiatry Clin. Neurosci. 269, 949–962. https://doi.org/10.1007/s00406-018-0943-x

LeCun, Y., Bottou, L., Bengio, Y., Haffner, P., 1998. Gradient-based learning applied to document recognition. Proc. IEEE 86, 2278–2323. https://doi.org/10.1109/5.726791

Ludwig, F.C., Smoke, M.E., 1980. The measurement of biological age. Exp. Aging Res. 6, 497–522. https://doi.org/10.1080/03610738008258384

Pedregosa, F., Varoquaux, G., Gramfort, A., Michel, V., Thirion, B., Grisel, O., Blondel, M., Prettenhofer, P., Weiss, R., Dubourg, V., Vanderplas, J., Passos, A., Cournapeau, D., Brucher, M., Perrot, M., Duchesnay, É., 2011. Scikit-learn: Machine learning in Python. J. Mach. Learn. Res.

Ronneberger, O., Fischer, P., Brox, T., 2015. U-net: Convolutional networks for biomedical image segmentation. Lect. Notes Comput. Sci. (including Subser. Lect. Notes Artif. Intell. Lect. Notes Bioinformatics) 9351, 234–241. https://doi.org/10.1007/978-3-319-24574-4_28

Smith, L.N., 2018. A disciplined approach to neural network hyper-parameters: Part 1 -- learning rate, batch size, momentum, and weight decay.

Smith, S.M., Vidaurre, D., Alfaro-Almagro, F., Nichols, T.E., Miller, K.L., 2019. Estimation of brain age delta from brain imaging. Neuroimage 200, 528–539. https://doi.org/10.1016/j.neuroimage.2019.06.017

Sokol, K., Santos-Rodriguez, R., Flach, P., 2019. FAT Forensics: A Python Toolbox for Algorithmic Fairness, Accountability and Transparency.

Srivastava, N., Hinton, G., Krizhevsky, A., Sutskever, I., Salakhutdinov, R., 2014. Dropout: A simple way to prevent neural networks from overfitting. J. Mach. Learn. Res.

Teuber, A., Sundermann, B., Kugel, H., Schwindt, W., Heindel, W., Minnerup, J., Dannlowski, U.,

Berger, K., Wersching, H., 2017. MR imaging of the brain in large cohort studies: feasibility report of the population- and patient-based BiDirect study. Eur. Radiol. https://doi.org/10.1007/s00330-016-4303-9

Vogelbacher, C., Möbius, T.W.D., Sommer, J., Schuster, V., Dannlowski, U., Kircher, T., Dempfle, A., Jansen, A., Bopp, M.H.A., 2018. The Marburg-Münster Affective Disorders Cohort Study (MACS): A quality assurance protocol for MR neuroimaging data. Neuroimage 172, 450–460. https://doi.org/10.1016/j.neuroimage.2018.01.079